\documentclass[aps,twocolumn,a4paper,floatfix,superscriptaddress,pra]{revtex4-1}
\usepackage[toc,page]{appendix}
\usepackage{braket}
\usepackage{epsfig,graphicx,times}
\usepackage{amstext}
\usepackage{amsmath}
\usepackage{amssymb}
\usepackage{mathrsfs}
\usepackage{dcolumn}
\usepackage{bm}
\usepackage[tight]{subfigure}
\usepackage[colorlinks,linkcolor=blue,anchorcolor=blue,citecolor=blue,urlcolor=blue]{hyperref}
\usepackage{color}
\usepackage{siunitx}
\usepackage{appendix}
\usepackage{placeins}
\usepackage{textcomp}
\usepackage{bbold}
\usepackage{float}
\usepackage{verbatim}
\usepackage{minitoc}
\usepackage[dvipsnames]{xcolor}

\usepackage{soul}
\usepackage[framemethod=tikz]{mdframed}
\usepackage{colortbl}  
\usepackage{xcolor}
\usepackage{array}

\begin{document}

\title{Asymmetry-enhanced phase sensing via asymmetric entangled coherent states}
\author{Xiao-Tong Chen}
\affiliation{Key Laboratory of Low-Dimensional Quantum Structures and Quantum Control of Ministry of Education, Department of Physics and Synergetic Innovation Center for Quantum Effects and Applications, Hunan Normal University, Changsha 410081, China}

\author{Wang-Jun Lu}\email{wjlu1227@zju.edu.cn}
\affiliation{Department of Maths and Physics, Hunan Institute of Engineering, Xiangtan 411104, China}

\author{Yunlan Zuo}
\affiliation{Key Laboratory of Low-Dimensional Quantum Structures and Quantum Control of Ministry of Education, Department of Physics and Synergetic Innovation Center for Quantum Effects and Applications, Hunan Normal University, Changsha 410081, China}

\author{Rui Zhang}
\affiliation{Key Laboratory of Low-Dimensional Quantum Structures and Quantum Control of Ministry of Education, Department of Physics and Synergetic Innovation Center for Quantum Effects and Applications, Hunan Normal University, Changsha 410081, China}

\author{Ya-Feng Jiao}\email{yfjiao@hunnu.edu.cn}
\affiliation{Key Laboratory of Low-Dimensional Quantum Structures and Quantum Control of Ministry of Education, Department of Physics and Synergetic Innovation Center for Quantum Effects and Applications, Hunan Normal University, Changsha 410081, China}

\author{Le-Man Kuang}\email{lmkuang@hunnu.edu.cn}
\affiliation{Key Laboratory of Low-Dimensional Quantum Structures and Quantum Control of Ministry of Education, Department of Physics and Synergetic Innovation Center for Quantum Effects and Applications, Hunan Normal University, Changsha 410081, China}
\affiliation{Academy for Quantum Science and Technology, Zhengzhou University of Light Industry, Zhengzhou 450002, China}

\date{\today}

\begin{abstract}
We study quantum phase sensing with an asymmetric two-mode entangled coherent state (ECS) in which the
two local amplitudes have different values. We find the phenomenon of the asymmetry-enhanced phase sensing
which the asymmetry can significantly increase the precise of the phase estimation. We further study the effect
of decoherence induced by the photon loss on quantum phase sensing. It is shown that the asymmetric ECSs
have stronger capability against decoherence over the symmetric ECSs. It is indicated that the asymmetric
ECSs have obvious advantages over the symmetric ECSs in the quantum phase sensing. We also study the
practical phase sensing scheme with the intensity-difference measurement, and show that the asymmetry in the
asymmetric ECSs can enhance the phase sensitivity in the practical phase measurement scheme. Our work
reveals the asymmetry in the asymmetric ECSs is a new quantum-sensing resource, and opens a new way to the
ultra-sensitive quantum phase sensing in the presence of photon losses.
\end{abstract}

\maketitle

\section{\label{level}Introduction}

Entangled coherent states (ECSs)~\cite{1,2,3,4,5} as a significant quantum resource of continuous variables have many potential applications in fundamental tests of quantum physics and
quantum information processing such as Bell-inequality tests~\cite{6,7,8}, tests for nonlocal realism~\cite{9,10}, quantum teleportation~\cite{11,12,13,14,15}, quantum computation~\cite{16,17,18,19,20,21,22}, quantum key distribution~\cite{23}, and precision measurements~\cite{24,25,26,27,28,29,30,31}. An optical four-component ECS~\cite{32} was experimentally prepared by using of a very lossy quantum channel. A two-mode ECS
was experimentally realized by using a Mach-Zehnder interferometer (MZI) equipped with a cross-Kerr element in each of two
spatially separated modes~\cite{33}. A number of schemes for implementation of ECSs in various quantum systems have been proposed~\cite{34,35,36,37,38,39,40,41,42,43,44,45,46,47}.

Enhancing the precision of a measured parameter is always the basic topics  in quantum metrology~\cite{48,49,50,51,52,53,54}.
Quantum mechanics imposes limits to the measurement precision.  Conventional measurement techniques typically fail to reach these quantum limits. Conventional bounds to the precision of measurements
such as  the standard quantum limit (SQL) are not as fundamental as the Heisenberg limits from the Heisenberg uncertainty principle, and can be beaten by using quantum strategies.
Quantum-enhanced metrology studies how to exploit quantum resources, such as squeezing, entanglement, and quantum phase transition to overcome the SQL and to exhibit quantum advantages~\cite{55,56,57,58,59,60,61,62,63,64,65,66,67,68,69,70,71,72,73,74,75,76}.


Phase estimation is a ubiquitous measurement primitive, used for precision measurement of length, displacement, speed, optical properties, and so on. Precise phase estimation is of particular significance for various applications such as imaging, sensing, and information processing. Caves~\cite{77} proposed  the first squeezing-enhanced MZI scheme for phase sensing  by taking a high-intensity coherent state and a low-intensity squeezed vacuum state as the input states of the MZI, which shows that the precision of phase estimation  can beat the SQL. This principle is widely used in gravitational wave observatories, with squeezed vacuum used to enhance precision beyond the limits of classical technology~\cite{78,79}.
Since Caves scheme of  quantum phase sensing, many protocols have been proposed to improve the precision of phase estimation, such as NOON states~\cite{80,81,82}, entangled coherent states~\cite{83}, two-mode squeezed states~\cite{84}, number squeezed states~\cite{85}, and so on.

A symmetric ECS consisting of a specific coherent superposition of the NOON states was proposed as the input state of the MZI phase measurement scheme for enhanced precision of phase sensing~\cite{83,86,87}. It was shown that  the precision of the NOON-type ECS can surpass that of the NOON state in both of the absence and presence of photon losses. Recently, it has been demonstrated that the symmetric ECS $|\alpha,\beta\rangle + |\beta,\alpha\rangle$ is the optimal ECS in lossy quantum-enhanced metrology for the MZI phase sensing scheme. In this paper, we  study how to effectively enhance the phase sensitivity  for quantum phase sensing by using asymmetric ECSs $|\alpha,\beta\rangle + |-\alpha, -\beta\rangle$.
We find that the asymmetry of asymmetric ECSs  can significantly enhance the precise of the phase sensing. We show that the asymmetric ECSs
have stronger capability against decoherence over the symmetric ECSs. It is indicated that the asymmetric
ECSs have obvious advantages over the symmetric ECSs in the quantum phase sensing. We also study the
practical phase sensing scheme with the intensity-difference measurement, and show that the asymmetry in the
asymmetric ECSs can enhance the phase sensitivity in the practical phase measurement scheme.

This paper is structured as follows. In Sec.\,\ref{level2}, we study the quantum phase sensing with asymmetric ECSs via the MZI in the absence of photon losses. In Sec.\,\ref{level3}, we investigate the quantum phase sensing with asymmetric ECSs under the effects of decoherence on the ECSs. In Sec.\,\ref{level4}, we discuss the quantum phase sensing in an optical intensity-difference-measurement scheme with asymmetric two-mode ECSs. Finally, our conclusions are summarized in Sec.\,\ref{level5}.


\section{\label{level2} Asymmetry-enhanced phase sensing with asymmetric ECSs}

In this section, we study asymmetry-enhanced phase sensing with asymmetric ECSs in the absence of photon losses.
We will show that the performance of the asymmetric ECSs can surpass that of the symmetric ECSs in the quantum phase estimation.

\begin{figure}[t]
  \centering
  \includegraphics[width=0.45\textwidth]{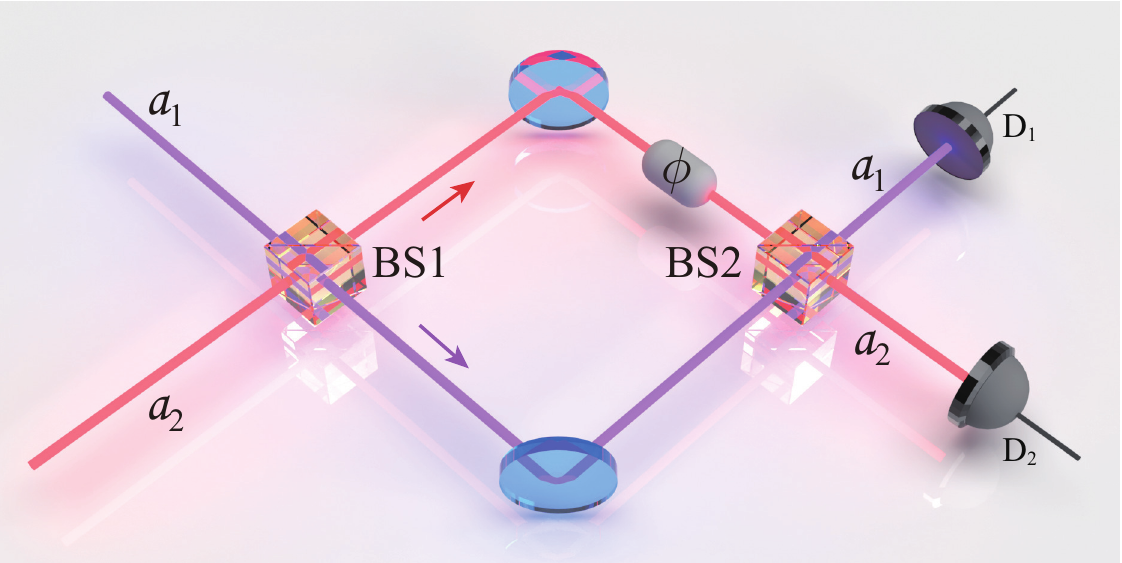}
  \caption{Schematic of the quantum phases sensing with the Mach-Zehnder interferometer.} \label{fig1}
\end{figure}

The MZI is a well-known optical device in quantum metrology which is constructed with two beam splitters and one or two phase shifts. We consider the MZI scheme of the quantum phases sensing~\cite{27} in which the MZI is constructed with two 50:50 beam splitters and one phase shift in one arm as shown in Fig.\,\ref{fig1}. We study the quantum phase sensing with the input sensor state after the first beam splitter being the following two-component ECS
\begin{equation}
\left|\Psi\right\rangle =N\left( \left|\alpha\right\rangle _{1}\left|k\alpha\right\rangle _{2} + \left|-\alpha\right\rangle _{1}\left|-k\alpha\right\rangle _{2}\right), \hspace{0.3cm} (k\neq 0),
\end{equation}
where $\left|\pm \alpha\right\rangle _{1}$  and  $\left|\pm k\alpha\right\rangle _{2}$ are coherent states with amplitudes $\pm \alpha$ and $\pm k\alpha$ for the first and second
sensor mode, respectively. $k$ is an arbitrary nonzero real number. The normalization constant is given by
\begin{equation}
N=\left[2 + 2  e^{-2(1+k^2)|\alpha|^{2}}\right]^{-1/2}.
\end{equation}

We can see that the ECSs given by Eq.\,(1) are symmetric or anti-symmetric under the exchange of the two sensor modes when $k=1$ and $k=-1$. They are regarded as symmetric and anti-symmetric ECS, respectively. When $k\neq \pm 1$, we call $\left|\Psi\right\rangle$ the asymmetric ECS. In what follows we will show that the asymmetry of  the asymmetric ECS would be a quantum resource to enhance quantum advantage for the quantum phase sensing.

In order to investigate achievable phase sensitivity with the asymmetric ECSs, we directly evaluate the quantum Fisher information (QFI) of
the output state of the MZI sensor after phase accumulation with the following expression
\begin{align}
\left|\Psi\left(\phi\right)\right\rangle &= e^{i\phi\hat{n}_{2}}\left|\Psi\right\rangle  \nonumber\\
&=N\left[\left|\alpha\right\rangle _{1}\left|e^{i\phi}k\alpha\right\rangle _{2} + \left|-\alpha\right\rangle _{1}\left|-e^{i\phi}k\alpha\right\rangle _{2}\right],
\end{align}
where $\hat{n}_{2}=\hat{a}_{2}^{\dagger}\hat{a}_{2}$ is the photon number operator of the second sensor mode.

The QFI of the  output state of the phase sensor  $\left|\Psi\left(\phi\right)\right\rangle$ with respect to the phase parameter $\phi$ is given by
\begin{align}
F_{Q}&=4\left(\left\langle \Psi^{\prime}\left(\phi\right)\right|\left.\Psi^{\prime}\left(\phi\right)\right\rangle -\left|\left\langle \Psi^{\prime}\left(\phi\right)\right|\left.\Psi\left(\phi\right)\right\rangle \right|^{2}\right),
\end{align}
where the derivative is given by
\begin{align}
\left|\Psi^{\prime}\left(\phi\right)\right\rangle&=\frac{\partial}{\partial\phi}\left|\Psi\left(\phi\right)\right\rangle =i\hat{n}_{2}\left|\Psi\left(\phi\right)\right\rangle.
\end{align}
Then we can obtain
\begin{align}
F_{Q}&=4\left[\left\langle \hat{n}^2_{2} \right\rangle -\left(\left\langle \hat{n}_2\right\rangle \right)^{2}\right],
\end{align}
where the expectation values are taken with respect to the sensor input state $\left|\Psi\right\rangle$ given by Eq.\,(1).

It is noted that the phase sensitivity should be determined by a relation between  the QFI  and the total mean photon number in the input state of the two sensor modes  $\left|\Psi\right\rangle$. The total photon number operator is $\hat{n}=\hat{n}_1+\hat{n}_2$ with $\hat{n}_{1}=\hat{a}_{1}^{\dagger}\hat{a}_{1}$ being the photon number operator of the first sensor mode. It is straightforward to obtain the following  the expectation values of the photon number operators in the input state $\left|\Psi\right\rangle$ given by Eq.\,(1), i.e.,
\begin{align}
\left\langle \hat{n}_{1}\right\rangle&=\left|\alpha\right|^{2} \frac{1 - e^{-2\left|\alpha\right|^{2}\left(1+k^{2}\right)}}{1 + e^{-2\left|\alpha\right|^{2}\left(1+k^{2}\right)}},\\
\left\langle \hat{n}_{2}\right\rangle&=k^{2}\left|\alpha\right|^{2}\frac{1 - e^{-2\left|\alpha\right|^{2}\left(1+k^{2}\right)}}{1 + e^{-2\left|\alpha\right|^{2}\left(1+k^{2}\right)}}, \\
\left\langle \hat{n}\right\rangle&=\left(1+k^{2}\right)\left|\alpha\right|^{2}\frac{1 - e^{-2\left|\alpha\right|^{2}\left(1+k^{2}\right)}}{1 + e^{-2\left|\alpha\right|^{2}\left(1+k^{2}\right)}}.
\end{align}
Then we have
\begin{equation}
\langle \hat{n}_{1}\rangle=\frac{1}{1+k^{2}}\left\langle \hat{n}\right\rangle, \hspace{0.5cm} \left\langle \hat{n}_{2}\right\rangle=\frac{k^{2}}{1+k^{2}}\left\langle \hat{n}\right\rangle.
\end{equation}
Similarly, we can get the following expectation values
\begin{align}
\left\langle \hat{n}_{2}^{2}\right\rangle&=\frac{k^{2}}{1+k^{2}}\left\langle \hat{n}\right\rangle +k^{4}\left|\alpha\right|^{4}, \\
\left\langle \hat{n}_{1}^{2}\right\rangle&=\frac{1}{1+k^{2}}\left\langle \hat{n}\right\rangle +\left|\alpha\right|^{4}, \\
\left\langle \hat{n}^{2}\right\rangle&=\left\langle \hat{n}\right\rangle +\left(1+k^{2}\right)^{2}\left|\alpha\right|^{4}.
\end{align}

Substituting Eqs.\,(10)-(13) into Eq.\,(6), we can express the QFI of the output state of the MZI sensor $\left|\Psi\left(\phi\right)\right\rangle$ in terms of the total mean photon number and its covariance as
\begin{align}
F_{Q}&=\left(\frac{2k}{1+k^{2}}\right)^2\left[ \left\langle \hat{n}\right\rangle+k^2(\Delta\hat{n})^2\right],
\end{align}
where $\left\langle \hat{n}\right\rangle$ and $(\Delta \hat{n})^2$ are the expectation value and the covariance of the total photon number operator in the input state of the phase sensor $\left|\Psi\right\rangle$ given by Eq.\,(1).

It is interesting to note that in the large  asymmetric regime of $k\gg 1$ from the expression of the Eq.(14) we can find that the ultimate precision of the phase sensitivity obeys the following inequality
\begin{equation}
\delta\phi_{min} < \frac{k}{2\sqrt{\left\langle \hat{n}\right\rangle+k^2(\Delta \hat{n})^2}},
\end{equation}
which indicates that the  phase sensitivity limit can surpass the standard quantum limit (SQL) and attain  the sub-Heisenberg and even Heisenberg  limit which depends upon the value of $(\Delta \hat{n})^2$ and $k$.

\begin{figure}[t]
  \centering
  \includegraphics[width=0.40\textwidth]{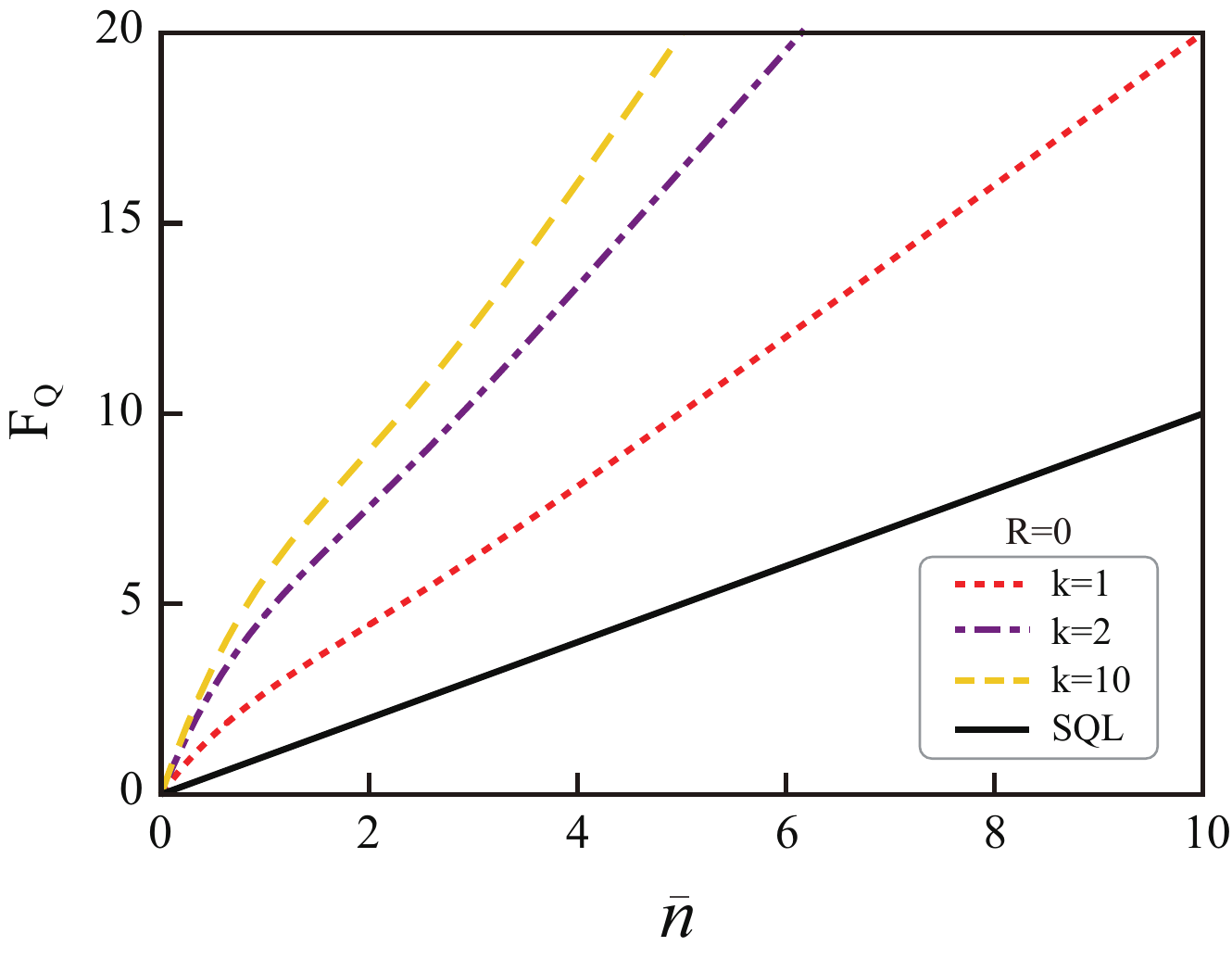}
  \caption{The QFI with respect to the total mean photon number $\bar{n}$.} \label{fig2}
\end{figure}

In Fig.\,\ref{fig2}, we plot the QFI with respect to the mean photon number $\bar{n}$ for different values of the asymmetric parameter. The dot, dashed-dot, and dashed lines correspond to $k=1, 2,$ and $10$, respectively. The solid line denotes SQL. From Fig.\,\ref{fig2}, we can observe that the QFI in the whole regime of $k>1$ is larger than the QFI of the SQL. The QFI of asymmetric ECSs is larger than that of the symmetric ECSs for the same total mean photon number $\bar{n}$. This implies that the phase sensitivity limit with asymmetric ECSs is lower than that with symmetric ECSs. In particular, Fig.\,\ref{fig2} indicates that in the small mean-photon-number regime about $\bar{n}<3$, we find $F_{Q}\sim \bar{n}^2$. So we have $\delta\phi_{min} \sim 1/\bar{n}$ due to $\delta\phi_{min} =1/\sqrt{F_Q}$. This means that the Heisenberg limit of the phase sensitivity is attainable  in the small mean-photon-number regime. Therefore, we can conclude that the asymmetry of the sensor input state can enhance the phase sensitivity limit.

In order to further explore the ultraprecise limit of the asymmetric ECSs in the phase estimation, in Table\,\ref{table1} we list some representative data about the QFI and the total mean photon number $\bar{n}$ and its square $\bar{n}^2$ when $k=10$. From data in Table\,\ref{table1}, we can see that $F_Q>\bar{n}^2$, i.e., $\delta\phi_{min} <1/\bar{n}$, when $1<\bar{n}\leq4$. This implies that the phase sensitivity limit of the asymmetric ECSs can surpass the Heisenberg limit in the small photon number. In the regime of $\bar{n}>4$, we can find $\bar{n}<F_Q<\bar{n}^2$, which indicates that the phase sensitivity limit of the asymmetric ECSs can surpass the SQL and attain the sub-Heisenberg limit.

\begin{table}
\begin{tabular}{|c|c|c|c|c|c|c|c|c|c|}
\hline
$\bar{n}$ & 1.0 & 1.5 & 2 & 2.5 & 3 & 3.5 & 4.0 & 4.5 & 5.0 \\
\hline
$\bar{n}^{2}$ & 1.00 & 2.25 & 4.00 & 6.25 & 9.00 & 12.25 & 16.00 & 20.25 & 25.00 \\
\hline
$F_Q$ & 5.78 & 7.44 & 9.05 & 10.48 & 12.36 & 13.97 & 16.05 & 17.98 & 20.54 \\
\hline
\end{tabular}
\caption{The QFI of the asymmetric ECSs with respect to the total mean photon number $\bar{n}$ and its square  $\bar{n}^2$  in the absence of photon losses. We here take the asymmetric parameter $k=10$.} \label{table1}
\end{table}

\section{\label{level3} Quantum phase sensing under decoherence}

In this section, we study the quantum phase sensing under the decoherence induced by photon losses. We shall calculate the QFI with decoherence and discuss the influence of the decoherence on the ultimate precision of the phase sensitivity.
The photon losses can be modeled by inserting two identical beam splitters in each optical arm of the MZI phase sensor  with the following beam splitter transformation
\begin{equation}
\hat{B}_{k,k'}(\gamma)=\exp\left[i(\gamma/2)(\hat{a}_{k'}^{\dagger}\hat{a}_{k}+ \hat{a}_{k}^{\dagger}\hat{a}_{k'})\right],
\end{equation}
which couple two optical sensor modes $k = 1, 2 $ and two environment modes  $k'= 1, 2 $ that are initially in the vacuum. The beam splitters transform the optical sensor mode into a linear combination of the optical sensor modes and  environment modes
\begin{equation}
\hat{B}_{k,k'}(\gamma) \hat{a}_{k}^{\dagger} \hat{B}^{-1}_{k,k'}(\gamma)=\sqrt{T}\hat{a}_{k}^{\dagger} + i\sqrt{R}\hat{a}_{k'}^{\dagger},
\end{equation}
where $T = \cos^2(\gamma/2)$ and $R = 1-T$ are transmission and  loss  rates of the photons, respectively.  When $T = 1$ ($R =0$), there are no photon losses in the interferometer, and when
$T = 0$ ($R = 1$), all the photons leak out of the interferometer.

We consider the case in which the input sensor state is $\left|\Psi\right\rangle$ given by Eq.\,(1) while the two environment modes are in the vacuum state. We assume that the leaks in both arms share the same transmission coefficient $T$. Then the total output state of the sensor modes and environment modes before the second beam splitter is given by
\begin{align}
\left|\Psi(\phi)\right\rangle&=\hat{B}_{1,1'}(\gamma)\hat{B}_{2,2'}(\gamma)\hat{U}\left(\phi\right)\left|\Psi _{+}\right\rangle|0, 0\rangle_{1',2'},
\end{align}
where the phase shift transformation is defined by $\hat{U}\left(\phi\right)=e^{i\phi\hat{n}_2}$.

After the actions of the beam-splitter and phase-shift transformations, Eq.\,(18) becomes
\begin{align}
|\Psi(\phi)\rangle&=N \left[|\sqrt{T}\alpha\rangle_{1}|i\sqrt{R}\alpha\rangle_{1'} \right. \nonumber\\
&\times \left.|e^{i\phi}\sqrt{T}k\alpha\rangle_{2}|e^{i\phi}i\sqrt{R}k\alpha\rangle_{2'} \right.\nonumber\\
&\left. +|-\sqrt{T}\alpha\rangle_{1}|-i\sqrt{R}\alpha\rangle_{1'} \right. \nonumber\\
&\times \left.|-\sqrt{T}k\alpha e^{i\phi}\rangle_{2}|-i\sqrt{R}k\alpha e^{i\phi}\rangle_{2'}\right],
\end{align}
which can be simply expressed as
\begin{align}
\left|\Psi(\phi)\right\rangle&=N_+\left[\left|\phi_1(T)\right\rangle  |i\sqrt{R}\alpha\rangle _{1^{'}}|e^{i\phi}i\sqrt{R}k\alpha\rangle _{2^{'}}\right.\nonumber\\
&\quad+\left.\left|\phi_2(T)\right\rangle|-i\sqrt{R}\alpha\rangle _{1^{'}}|-i\sqrt{R}k\alpha e^{i\phi}\rangle _{2^{'}}\right],
\end{align}
where we have introduced two sensor-mode states
\begin{align}
\left|\phi_{1}\left(T\right)\right\rangle&=|\sqrt{T}\alpha\rangle _{1}|e^{i\phi}\sqrt{T}k\alpha\rangle _{2},  \nonumber\\
\left|\phi_{2}\left(T\right)\right\rangle &=|-\sqrt{T}\alpha\rangle _{1}|-e^{i\phi}\sqrt{T}k\alpha\rangle_{2},
\end{align}
which have the following non-orthogonal relation
\begin{align}
\left\langle \phi_{1}\left(T\right)\right|\left.\phi_{2}\left(T\right)\right\rangle =e^{-2T(1+k^2)|\alpha|^{2}}.
\end{align}

The reduced density operator of the sensor modes is given by
\begin{align}
\hat{\rho}\left(T\right)&=N^{2}\left\{\left|\phi_{1}\left(T\right)\right\rangle \left\langle \phi_{1}\left(T\right)\right|+ \left|\phi_{2}\left(T\right)\right\rangle \left\langle \phi_{2}\left(T\right)\right|\right.\nonumber\\
&\quad+2 e^{-2R(1+k^2)|\alpha|^{2}}\left[\left|\phi_{1}\left(T\right)\right\rangle \left\langle \phi_{2}\left(T\right)\right|+\text{H.c.}\right]\}.
\end{align}
Making use of the diagonalization method of a density matrix developed in Ref.\,\cite{31}, through some tedious calculations we can obtain the eigenvalues and eigenfunctions of the reduced density operator of the sensor modes
\begin{align}
\lambda_{\pm}&=\frac{1}{2}(1\pm\Lambda), \\
\Lambda&=\sqrt{1-N^{4}\left(1-e^{-4T(1+k^2)|\alpha|^{2}}\right)(1-e^{-4R(1+k^2)|\alpha|^{2}})}.
\end{align}

And two orthogonal eigenfunctions of the reduced density operator are given by
\begin{align}
\left|\lambda_{\pm}\left(\phi\right)\right\rangle&=M_{\pm}(\eta_{\pm}|\sqrt{T}\alpha\rangle _{1}|e^{i\phi}\sqrt{T}k\alpha\rangle _{2}  \nonumber\\
&+|-\sqrt{T}\alpha\rangle _{1}|-e^{i\phi}\sqrt{T}k\alpha\rangle _{2}),
\end{align}
where the normalization constants are given by
\begin{align}
M_{\pm}&=\left[\left(1+\eta^2_{\pm}\right)+ 2\eta_{\pm}e^{-2T(1+k^2)|\alpha|^{2}}\right]^{-1/2}, \\
\eta_{\pm}&=\frac{N^{2}\left[2e^{-2T(1+k^2)|\alpha|^{2}} + e^{-2R(1+k^2)|\alpha|^{2}}\right]}{2\lambda_{\pm}-N^{2}\left[2+ e^{-2(1+k^2)|\alpha|^{2}}\right]}.
\end{align}

Since the eigenvalues of the reduced density operator of the sensor modes are independent of the phase parameter $\phi$, we can calculate the QFI  by the use of the following formula \cite{31}
\begin{align}
	F_{Q}&=\sum_{i=\pm}\lambda_{i}F_{Q,i}-\sum_{i\neq j}\frac{8\lambda_{i}\lambda_{i}}{\lambda_{i}+\lambda_{j}}\left|\left\langle \lambda_{i}^{\prime}\right|\left.\lambda_{j}\right\rangle \right|^{2},
\end{align}
where the first term is the QFI of the eigenvalues  of the reduced density operator of the sensor modes
\begin{equation}
F_{Q \pm}=4(\langle\lambda_{\pm}^{\prime}|\lambda_{\pm}^{\prime}\rangle-|\langle\lambda_{\pm}^{\prime}|\lambda_{\pm}\rangle |^2 ),
\end{equation}

For convenience of calculation, we rewrite Eq.\,(29) as
\begin{align}
F_{Q}&=\left[\left(F_{Q1,+} -F_{Q2,+}^{\text{}}\right)+\left(F_{Q1,-}-F_{Q2,-}^{\text{}}\right)\right] \nonumber\\
&-\left(F_{Q3,+}+F_{Q3,-}\right),
\end{align}
where we introduce the following functions
\begin{align}
F_{Q1,\pm} &=4\lambda_{\pm}\left\langle \lambda_{\pm}^{\prime}|\lambda_{\pm}^{\prime}\right\rangle, \\
F_{Q2,\pm}&= 4\lambda_{\pm}|\left\langle \lambda_{\pm}^{\prime}|\lambda_{\pm}\right\rangle|^2,  \\
F_{Q3,\pm}&=\frac{8\lambda_{+}\lambda_{-}}{\lambda_{+}+\lambda_{-}}\left|\left\langle \lambda_{\pm}^{\prime}\right|\left.\lambda_{\mp}\right\rangle \right|^{2}.
\end{align}

Making use of the eigenvalues and  eigenfunctions of the reduced density operator given by Eqs.\,(24) and (26) it is straightforward to obtain
\begin{align}
F_{Q1,\pm}&=4\lambda_{\pm}\left|M_{\pm}\right|^{2}Tk^2\left|\alpha\right|^{2}\Bigl[\left(1+\eta^2_{\pm}\right)\left(1+Tk^2|\alpha|^{2}\right)\nonumber\\
&\quad-2\eta_{\pm}\left(1-Tk^2|\alpha|^{2}\right)e^{-2T(1+k^2)|\alpha|^{2}}\Bigl],\\
F_{Q2,\pm}&=4\lambda_{\pm}\left|M_{\pm}\right|^{4}T^{2}k^4|\alpha|^{4}\Bigl[\left(1+\eta^2_{\pm}\right)\nonumber \\
&-2\eta_{\pm}e^{-2T(1+k^2)|\alpha|^{2}}\Bigl]^{2}, \\
F_{Q3,\pm}&=\frac{8\lambda_{+}\lambda_{-}}{\lambda_{+}+\lambda_{-}}\left|M_{+}\right|^{2}\left|M_{-}\right|^{2}T^{2}k^4|\alpha|^{4}\nonumber\\
&\times \left|\left(1+\eta_{\pm}\eta_{\mp}\right)-e^{-2T(1+k^2)|\alpha|^{2}}\left(\eta_{\pm}+\eta_{\mp}\right)\right|^{2}.
\end{align}

\begin{figure}[t]
  \centering
  \includegraphics[width=0.47\textwidth]{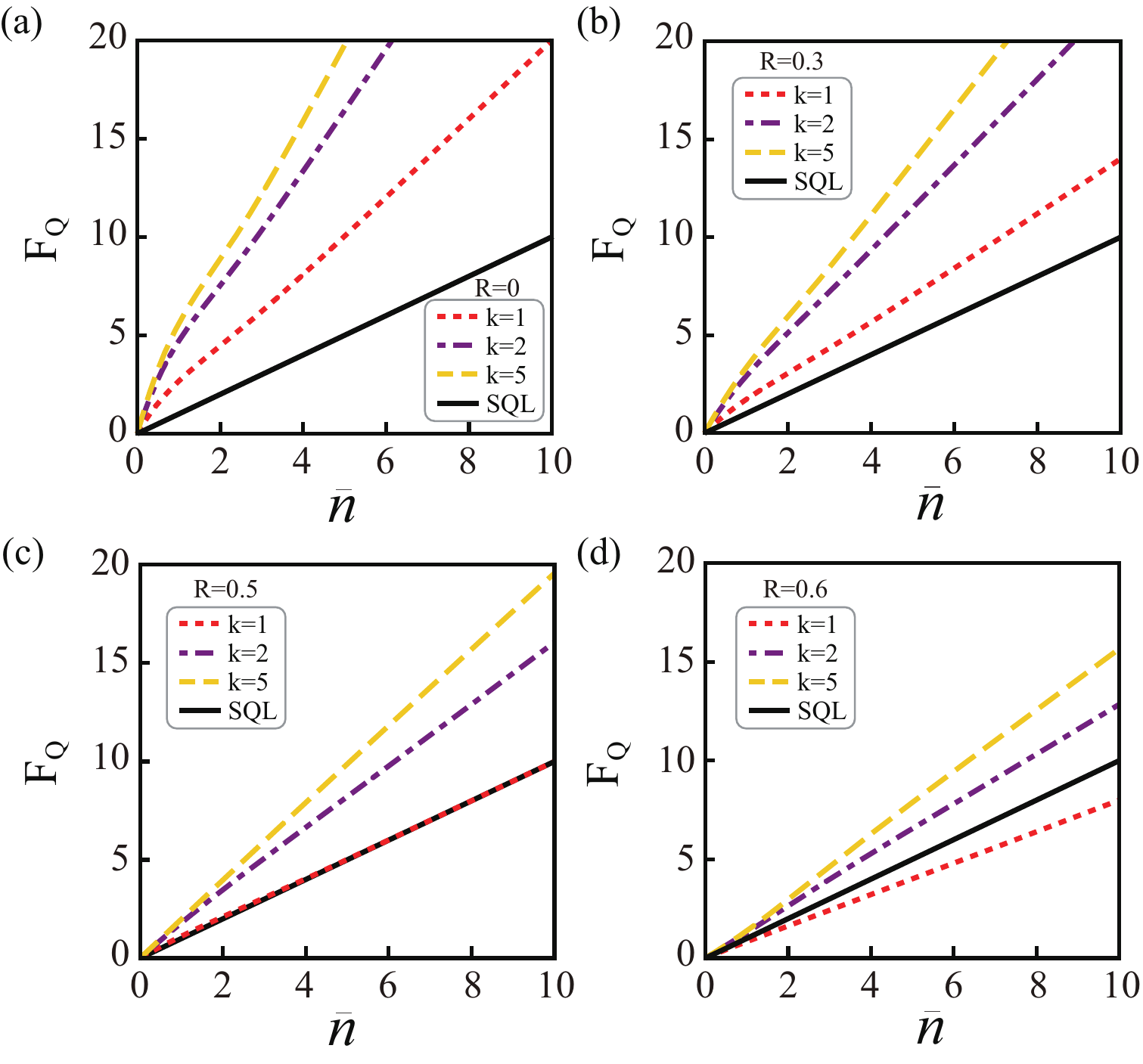}
  \caption{The QFI with respect to the total mean photon number for different values of the asymmetric parameter $k$.} \label{fig3}
\end{figure}

In what follows we  numerically analyze the influence of the asymmetry in the asymmetric ECSs on the phase sensitive limit determined by the QFI given by Eq.\,(31). In Fig.\,\ref{fig3}, we plot the QFI with respect to the total mean photon number for different values of the asymmetric parameter $k$ in the presence of photon losses.  The numerical simulation parameters are chosen as the photon loss rate $R=0, 0.3, 0.5, 0.6$, and the symmetric parameter $k= 1$ with asymmetric parameter  $k= 2$, $5$. The solid lines in Figs.\,\ref{fig3}(a)-\ref{fig3}(d) denote the QFI corresponding to the SQL. The dashed lines show the QFI in the case of the asymmetric ECSs.The dots lines show the QFI in the case of the symmetric ECSs.
Figures\,\ref{fig3}(a)-\ref{fig3}(d) show that in the symmetric-ECS case, the QFI decreases with the increase of the photon losses. The QFI is larger than that of the SQL in the regime of the smaller photon losses ($0<R<0.5$). The QFI is equal to that of the SQL when $R=0.5$. The QFI is smaller than that of the SQL in the regime of the larger photon losses ($R>0.5$).

However, the situation is significantly different in the presence of the state asymmetry.  The dashed-dot and dashed lines in Figs.\,\ref{fig3}(a)-\ref{fig3}(d) represent the QFI when the asymmetric parameter $k=2$ and $5$, respectively. We can observe that the QFI of the asymmetric ECSs is larger than that of the symmetric ECSs and the SQL even in the presence of strong photon losses with $R=0.6$ in Fig.\,\ref{fig3}(d).

\begin{table}
\begin{tabular}{|c|c|c|c|c|c|c|c|c|c|}
\hline
$\bar{n}$ & 1.0 & 1.5 & 2.0 & 2.5 & 3.0 & 3.5 & 4.0 & 4.5 & 5.0 \\
\hline
$\bar{n}^{2}$ & 1.00 & 2.25 & 4.00 & 6.25 & 9.00 & 12.25 & 16.00 & 20.25 & 25.00 \\
\hline
$F_Q$ & 2.92 & 4.06 & 5.11 & 6.15 & 7.21 & 8.26 & 9.36 & 10.40 & 11.50 \\
\hline
\end{tabular}
\caption{The QFI of the asymmetric ECSs with respect to the total mean photon number $\bar{n}$ and its square  $\bar{n}^2$ in the presence of photon losses. We here take the photon loss rate $R=0.3$ and the asymmetric parameter $k=2$. \label{table2}}
\end{table}

We can in detail observe the ultraprecise limit of the asymmetric ECSs in the phase estimation through comparing the QFI with values of $\bar{n}^2$. In Table\,\ref{table2}, we list some representative data about the QFI and the total mean photon number $\bar{n}$ and its square $\bar{n}^2$ when we take the photon loss rate $R=0.3$ and the asymmetric parameter $k=2$. From data in  Table\,\ref{table2}, we can see that $F_Q>\bar{n}^2$, i.e., $\delta\phi_{min} < 1/\bar{n}$, when $1\leq \bar{n}\leq 2$. This implies that the phase sensitivity limit of the asymmetric ECSs can surpass the Heisenberg limit in the small photon number. Especially, it can be shown that the phase sensitivity limit of the asymmetric ECSs can also surpass the Heisenberg limit in the small photon number even in the case of the strong photon losses. For instance, for the case of the strong photon loss with $R=0.6$ when $k=5$ we find that $F_Q>\bar{n}^2$, $\delta\phi_{min} < 1/\bar{n}$  is still valid in the small photon-number regime of $\bar{n}<1.5$.

\begin{figure}[t]
  \centering
  \includegraphics[width=0.40\textwidth]{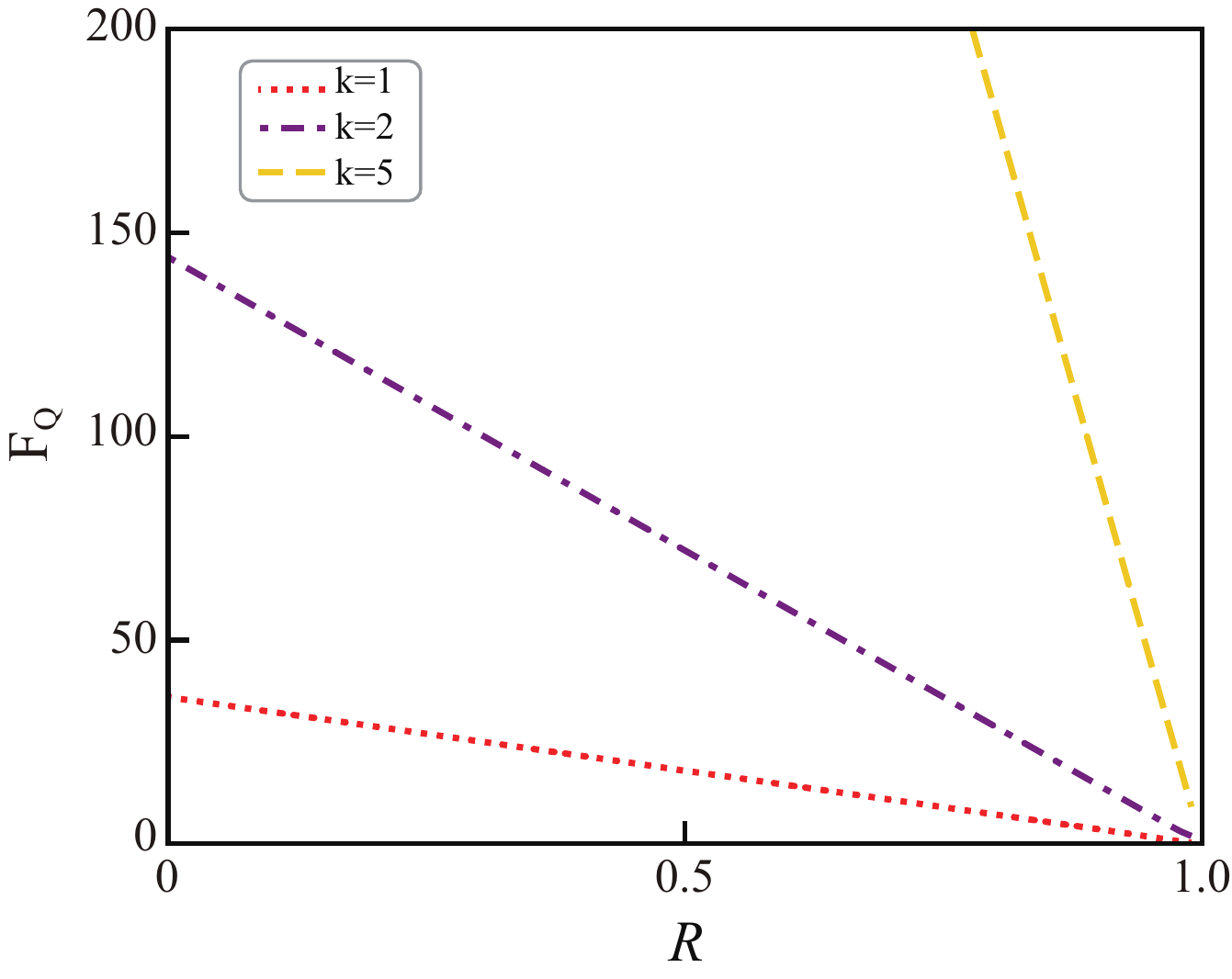}
  \caption{The QFI with respect to photon losses for different values of the asymmetric parameter $k$.} \label{fig4}
\end{figure}

In order to further demonstrate the asymmetric advantage of the asymmetric ECSs in the phase estimation, in Fig.\,\ref{fig4} we  numerically plot  the QFI with respect to photon losses for the symmetric and asymmetric ECSs where the asymmetric parameter takes  $k=2$, $5$,  respectively. The dots line is the symmetric-ECS case while the dashed-dot and dashed lines corresponds to the asymmetric-ECS cases with $k=2$ and $5$, respectively. From Fig.\,\ref{fig4} we can see that the QFI of the symmetric ECS rapidly decays with the photon losses while the QFI of the asymmetric ECSs decays much slower than the symmetric ECS. This implies that the the asymmetric ECSs have stronger capability against the photon losses. And the larger is the asymmetric degree of the asymmetric ECSs, the stronger is the capability  against the photon losses.

From the above numerical analysis we can conclude that  the asymmetric ECSs can exhibit the quantum advantage in the phase estimation over the symmetric ECSs in the presence of the photon losses. Firstly, the asymmetric ECSs can create better phase sensitivity limit than the symmetric ECSs. Secondly, the asymmetric ECSs have stronger capability against the photon losses. In particular, in the regime of the small photon number the phase sensitivity limit of the asymmetric ECSs can also surpass the Heisenberg limit  even in the environment with the strong photon losses.

\section{\label{level4} Quantum phase sensing with intensity-difference measurement}
In this section, we study the quantum phase sensing in a real
measurement scheme. We consider the ultimate phase sensitivity
with the MZI in the optical intensity-difference measurement
given by the following intensity-difference operator

\begin{align}
\hat{S}_{z}=\frac{1}{2}(\hat{a}^{\dagger}\hat{a}-\hat{b}^{\dagger}\hat{b}).
\end{align}

The sensor output state after the second beam splitter is given by
\begin{align}
\hat{\rho}_{\text{out}}\left(\phi\right)=\hat{B}_{1,2}\left(\pi/2\right)\hat{\rho}\left(\phi\right)\hat{B}_{1,2}^{\dagger}\left(\pi/2\right),
\end{align}
where $\hat{B}_{1,2}\left(\pi/2\right)$ is the second 50 : 50 beam splitter transformation in the MZI, and the reduced density matrix $\hat{\rho}\left(\phi\right)$ of the sensor
modes is given by Eq.\,(23) which can be expressed as
\begin{align}
\hat{\rho}\left(\phi\right)=\lambda_{+}\left|\lambda_{+}\left(\phi\right)\right\rangle \left\langle \lambda_{+}\left(\phi\right)\right|+\lambda_{-}\left|\lambda_{-}\left(\phi\right)\right\rangle \left\langle \lambda_{-}\left(\phi\right)\right.|.
\end{align}

It is straightforward to calculate the expectation value of the intensity-difference operator $\hat{S}_{z}$ in the output state $\hat{\rho}_{\text{out}}$ of the MZI
with the following result
\begin{align}
\langle\hat{S}_{z}\rangle =&\lambda_{+}\left\langle \lambda_{+}\right|\hat{B}_{1,2}^{\dagger}\hat{S}_{z}\hat{B}_{1,2}\left|\lambda_{+}\right\rangle +\lambda_{-}\left\langle \lambda_{-}\right|\hat{B}_{1,2}^{\dagger}\hat{S}_{z}\hat{B}_{1,2}\left|\lambda_{-}\right\rangle \nonumber\\
 =& S_{z,+}+S_{z,-},
\end{align}
where $S_{z,\pm}$ is given by
\begin{align}
S_{z,\pm} =&-\lambda_{\pm}kT\left|\alpha\right|^{2}\sin\phi M_{+}^{2}\Bigl[\left(1+\eta^2_{\pm}\right)\nonumber\\
 &-2\eta_{\pm}e^{-2T\left|\alpha\right|^{2}(k^{2}+1)}\Bigl].
\end{align}

Simmilarly, the expectation value of the operator   $\hat{S}^2_{z}$  in the sensor output state  $\hat{\rho}_{\text{out}}$ can be obtained with the following expression
\begin{align}
\langle\hat{S}^2_{z}\rangle& = S_{z2,+}+S_{z2,-},
\end{align}
where $S_{z2,\pm}$ is given by
\begin{align}
S_{z,\pm}  =&-\lambda_{\pm}kT\left|\alpha\right|^{2}\sin\phi M_{+}^{2}\Bigl[\left(1+\eta^2_{\pm}\right)\nonumber\\
 &-2\eta_{\pm}e^{-2T\left|\alpha\right|^{2}(k^{2}+1)}\Bigl].
\end{align}

Hence, the phase sensitivity in the intensity-difference measurement scheme is given by the error transfer equation
\begin{align}
\triangle\phi & =\frac{\sqrt{\langle \hat{S}_{z}^{2}\rangle -\langle \hat{S}_{z}\rangle^{2}}}{|\partial\langle \hat{S}_{z}\rangle /\partial\phi|}.
\end{align}

\begin{figure}[t]
  \centering
  \includegraphics[width=0.45\textwidth]{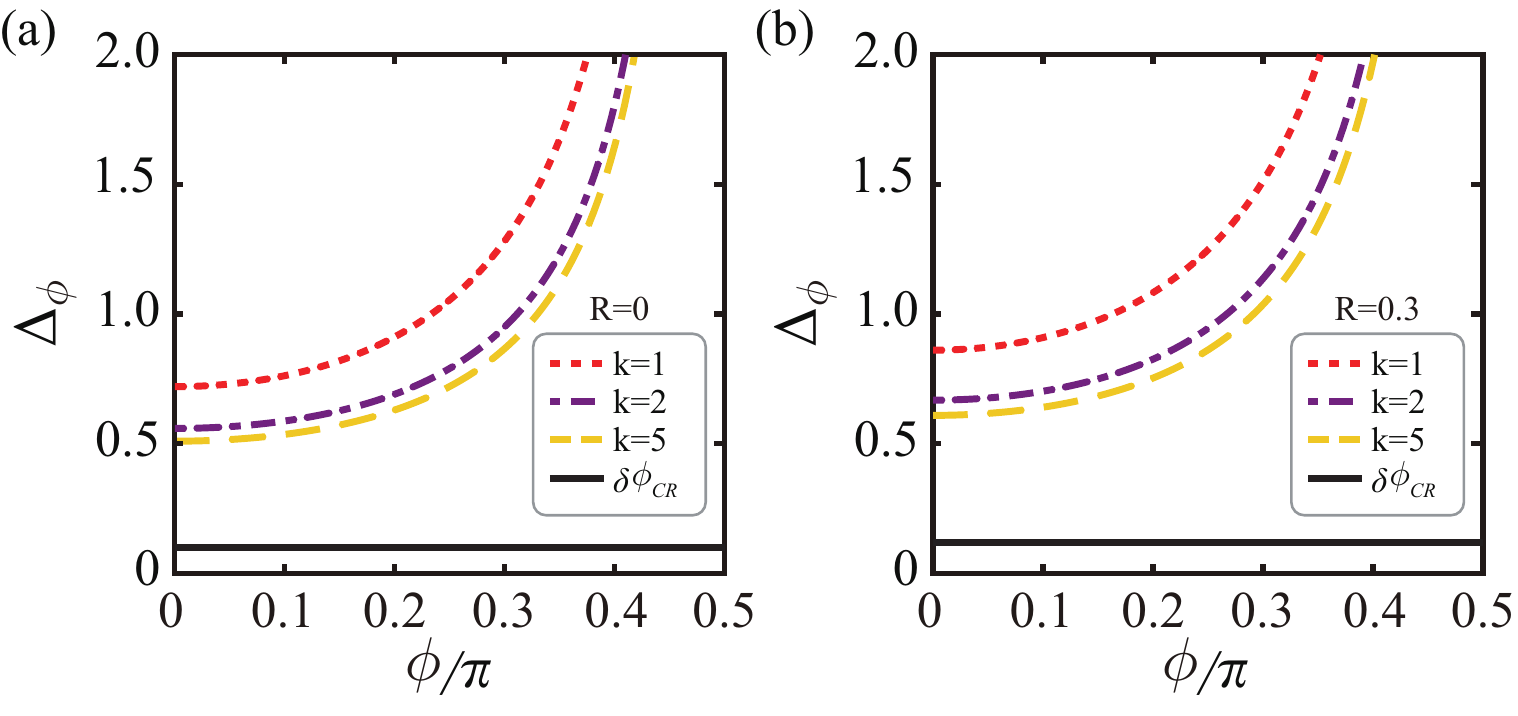}
  \caption{The phase sensitivity in the intensity-differencemeasurement
scheme for different values of the asymmetric parameter
$k$. (a) The case without photon losses. (b) The case in the presence
of photon losses.} \label{fig5}
\end{figure}

In Fig.\,\ref{fig5}, we plot the phase sensitivity in the intensity-difference-measurement scheme for different values of the asymmetric parameter $k$. Figures\,\ref{fig5}(a) and \ref{fig5}(b) correspond
to the two cases with and without photon losses, respectively. The solid line in Fig.\,\ref{fig5} denotes the Cram\'{e}r-Rao bound given by the previous section. From Fig.\,\ref{fig5} we can see that the phase sensitivity
described by $\triangle \phi$ becomes better with increasing the asymmetric parameter $k$ for both of the two cases without and with photon losses, and it more closely approaches to the Cram\'{e}r-Rao bound. Therefore, we can conclude that the asymmetry in the asymmetric ECS may enhance the phase sensitivity in the practical phase measurement scheme.

\section{\label{level5}Conclusions}

In this work, we have studied the quantum phase sensing with asymmetric two-mode ECSs via the MZI. We first investigate the quantum phase sensing with asymmetric ECSs in the absence of photon losses. We have shown that the asymmetry of the input asymmetric ECS can be used as a quantum resource to enhance the sensitivity of the quantum phase estimation. In particularly, we have found that the phase sensitivity limit of the asymmetric ECSs can attain and even surpass the Heisenberg limit in certain parameter regimes.

We then studied the quantum phase sensing with asymmetric two-mode ECSs under the effects of decoherence on the
ECSs. We use an optical beam splitter to describe the photon losses. We have indicated that the asymmetric ECSs can exhibit
the quantum advantage in the phase estimation over the symmetric ECSs in the presence of the photon losses. We have
found that the asymmetric ECSs can not only create the better phase sensitivity limit than the symmetric ECSs, but also have
stronger capability against the photon losses. In particular, in the regime of the small photon number the phase sensitivity
limit of the asymmetric ECSs can also surpass the Heisenberg limit even in the environment with the strong photon losses.

We have also investigated the quantum phase sensing in a real measurement scheme with asymmetric two-mode ECSs.
We have studied the ultimate phase sensitivity with the MZI in the optical intensity-difference-measurement scheme. We
found that the asymmetry in the asymmetric ECSs can also enhance the phase sensitivity in the practical phase measurement
scheme. In summary, Our results reveal the asymmetry in the asymmetric ECSs as a new quantum-sensing resource
and open a new way to the ultra-sensitive quantum phase sensing in the presence of photon loss.

\begin{acknowledgements}
L.-M. K. is supported by the Natural Science Foundation of China (NSFC) Grant Nos. 12247105, 1217050862, and 11935006. W.-J. L. is supported by the NSFC (Grant No. 12205092). Y.-F. J. is supported by the NSFC (Grant No. 12147156), the China Postdoctoral Science Foundation (Grants No. 2021M701176 and No. 2022T150208), and the Science and Technology Innovation Program of Hunan Province (Grant No. 2021RC2078).
\end{acknowledgements}

%

%

\end{document}